%% file: main.tex
\PassOptionsToPackage{table}{xcolor}
\documentclass[acmtog,screen,nonacm]{acmart}

\AtBeginDocument{%
  \providecommand\BibTeX{{%
    \normalfont B\kern-0.5em{\scshape i\kern-0.25em b}\kern-0.8em\TeX}}}

\setcopyright{acmlicensed}
\copyrightyear{2024}
\acmYear{2024}

\usepackage{adjustbox}
\usepackage{bm}
\usepackage[ruled]{algorithm2e} 

\SetAlFnt{\small}
\SetAlCapFnt{\small}
\SetAlCapNameFnt{\small}
\SetAlCapHSkip{0pt}
\usepackage{xcolor}
\usepackage{wrapfig}
\usepackage{mathtools}
\usepackage{units} 
\usepackage{amsmath}
\usepackage{overpic}
\usepackage{color, colortbl}
\usepackage{tikz}
\usepackage{enumitem}
\usepackage{listings}

\colorlet{punct}{red!60!black}
\definecolor{background}{HTML}{EEEEEE}
\definecolor{delim}{RGB}{20,105,176}
\colorlet{numb}{magenta!60!black}

\lstdefinelanguage{json}{
    basicstyle=\scriptsize\ttfamily,
    numbers=none,
    numberstyle=\scriptsize,
    stepnumber=1,
    numbersep=8pt,
    showstringspaces=false,
    breaklines=true,
    frame=single,
    backgroundcolor=\color{background},
    literate=
     *{0}{{{\color{numb}0}}}{1}
      {1}{{{\color{numb}1}}}{1}
      {2}{{{\color{numb}2}}}{1}
      {3}{{{\color{numb}3}}}{1}
      {4}{{{\color{numb}4}}}{1}
      {5}{{{\color{numb}5}}}{1}
      {6}{{{\color{numb}6}}}{1}
      {7}{{{\color{numb}7}}}{1}
      {8}{{{\color{numb}8}}}{1}
      {9}{{{\color{numb}9}}}{1}
      {:}{{{\color{punct}{:}}}}{1}
      {,}{{{\color{punct}{,}}}}{1}
      {\{}{{{\color{delim}{\{}}}}{1}
      {\}}{{{\color{delim}{\}}}}}{1}
      {[}{{{\color{delim}{[}}}}{1}
      {]}{{{\color{delim}{]}}}}{1},
}

\definecolor{turquoise}{cmyk}{0.65,0,0.1,0.1}
\definecolor{purple}{rgb}{0.65,0,0.65}
\definecolor{darkgreen}{rgb}{0.0, 0.5, 0.0}
\definecolor{darkred}{rgb}{0.5, 0.0, 0.0}
\definecolor{darkblue}{rgb}{0.0, 0.0, 0.5}
\definecolor{blue}{rgb}{0.0, 0.0, 1.0}
\definecolor{orange}{rgb}{1.0, 0.5, 0.0}
\definecolor{red}{rgb}{1.0, 0.0, 0.0}
\definecolor{cherry}{RGB}{186,12,47}
\definecolor{lightgray}{gray}{0.9}

\newcommand{\refsec}[1] {Section~\ref{#1}}

\newcommand{\reffig}[1] {Figure~\ref{#1}}

\newcommand{\system}[0]{Script2Screen}
\newcommand{\user}[1]{\textit{#1}}

\newcommand{\userquote}[1]{\textit{``#1''}}
\newcommand{\dgone}[0]{\textbf{DG 1}}
\newcommand{\dgtwo}[0]{\textbf{DG 2}}
\newcommand{\dgthree}[0]{\textbf{DG 3}}
\newcommand{\dgfour}[0]{\textbf{DG 4}}

\begin{document}

\title{Script2Screen: Supporting Dialogue Scriptwriting with Interactive Audiovisual Generation}

\author{Zhecheng Wang}
\orcid{0000-0003-4989-6971}
\email{zhecheng@cs.toronto.edu}
\affiliation{%
  \institution{University of Toronto}
  \streetaddress{40 St. George Street}
  \city{Toronto}
  \state{ON}
  \country{Canada}
  \postcode{M5S 2E4}
}

\author{Jiaju Ma}
\orcid{0000-0003-2880-8506}
\email{jiajuma@stanford.edu}
\affiliation{%
  \institution{Stanford University}
  \streetaddress{353 Jane Stanford Way}
  \city{Stanford}
  \state{CA}
  \country{United States}
  \postcode{94305}
}

\author{Eitan Grinspun}
\orcid{0000-0003-4460-7747}
\email{eitan@cs.toronto.edu}
\affiliation{%
  \institution{University of Toronto}
  \streetaddress{40 St. George Street}
  \city{Toronto}
  \state{ON}
  \country{Canada}
  \postcode{M5S 2E4}
}

\author{Tovi Grossman}
\orcid{0000-0002-0494-5373}
\email{tovi@cs.toronto.edu}
\affiliation{%
  \institution{University of Toronto}
  \streetaddress{40 St. George Street}
  \city{Toronto}
  \state{ON}
  \country{Canada}
  \postcode{M5S 2E4}
}

\author{Bryan Wang}
\orcid{0000-0001-9016-038X}
\email{bryanw@adobe.com}
\affiliation{%
  \institution{Adobe Research}
  \streetaddress{801 N. 34th Street}
  \city{Seattle}
  \state{WA}
  \country{United States}
  \postcode{98103}
}

\renewcommand{\shortauthors}{Wang, et al.}

\begin{abstract}
Scriptwriting has traditionally been text-centric, a modality that only partially conveys the produced audiovisual experience. A formative study with professional writers informed us that connecting textual and audiovisual modalities can aid ideation and iteration, especially for writing dialogues. In this work, we present \system{}, an AI-assisted tool that integrates scriptwriting with audiovisual scene creation in a unified, synchronized workflow. Focusing on dialogues in scripts, \system{} generates expressive scenes with emotional speeches and animated characters through a novel text-to-audiovisual-scene pipeline. The user interface provides fine-grained controls, allowing writers to fine-tune audiovisual elements such as character gestures, speech emotions, and camera angles. A user study with both novice and professional writers from various domains demonstrated that \system{}’s interactive audiovisual generation enhances the scriptwriting process, facilitating iterative refinement while complementing - rather than replacing - their creative efforts.
\end{abstract}


\begin{teaserfigure}
  \includegraphics[width=\textwidth]{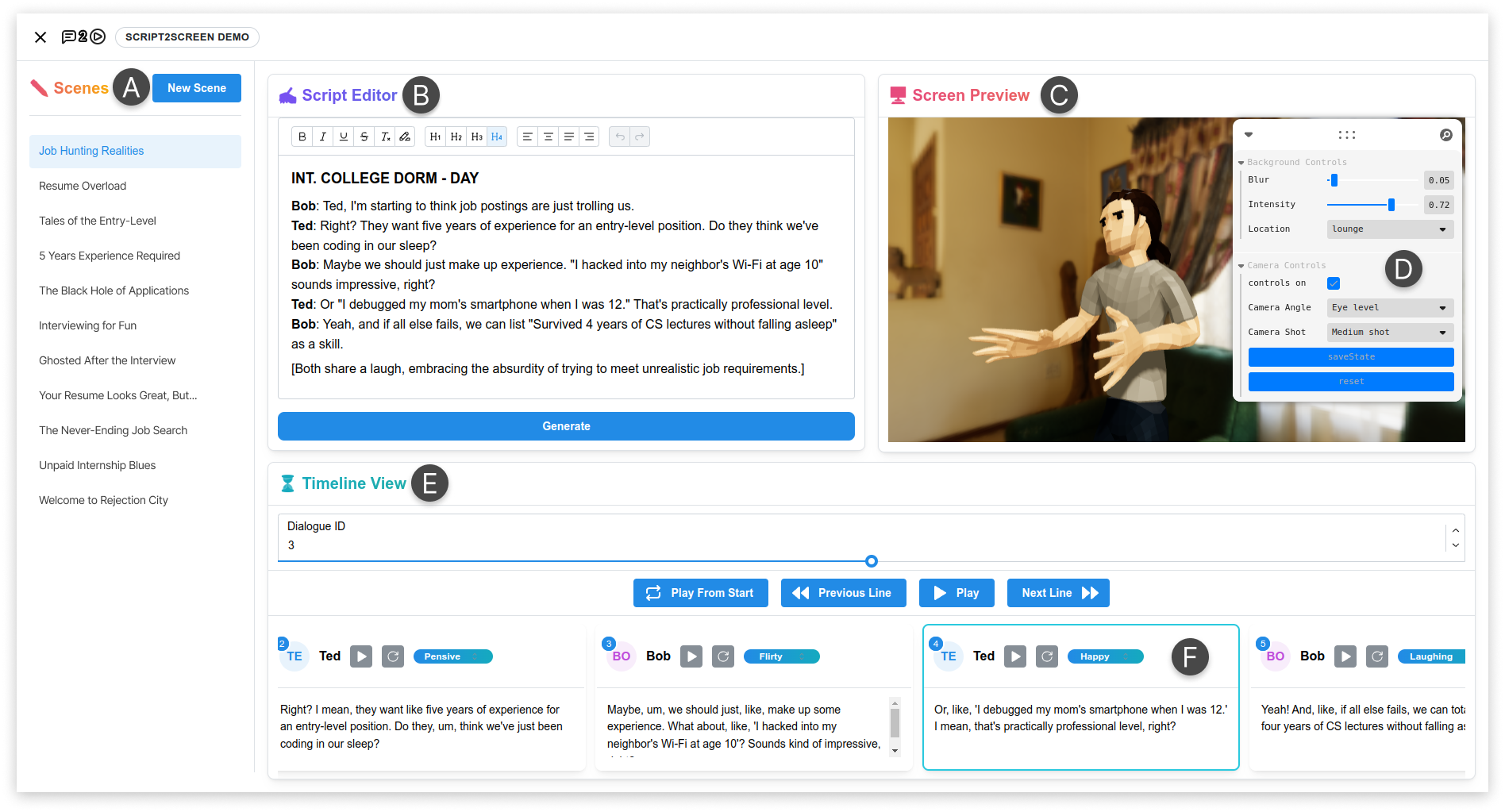}
  \caption{
    The \textit{\system{}} interface, captured from a writing session. \system{} allows screenwriters to seamlessly visualize and hear their scenes. Writers can select a workspace from the \textit{Scenes} navigation bar (A), where the corresponding script loads into the \textit{Script Editor} (B). After the script is processed, the \textit{Screen Preview} (C) offers a live 3D animation with synced audio, complete with customizable camera angles and background settings (D). The \textit{Timeline View} (E) provides a streamlined, interactive breakdown of each dialogue segment, showcasing emotions, speaker cues, and playback controls in dialogue cards (F). \system{} delivers a fully immersive WYSIWYG experience, making it an ideal tool for crafting scripts, brainstorming ideas, or generating dynamic storyboards ready for downstream applications.}
  \Description{Pipeline of Script2Screen.}
  \label{fig:teaser}
\end{teaserfigure}

\maketitle

\input{sections/introduction.tex}
\input{sections/related_works.tex}
\input{sections/formative_studies.tex}
\input{sections/design_objectives.tex}
\input{sections/user_interface}
\input{sections/backend.tex}

\input{sections/user_studies.tex}
\input{sections/results}
\input{sections/discussion.tex}
\input{sections/conclusion.tex}
\input{sections/genaidisclosure.tex}

\bibliographystyle{ACM-Reference-Format}
\bibliography{main}

\appendix
\input{sections/appendix.tex}


\end{document}

%% file: sections/introduction.tex
\section{Introduction}
\label{sec:introduction}
Scriptwriting transforms ideas into text, forming the basis for audiovisual content such as films and animations.
When developing stories, writers frequently use dialogues to build narratives, shape characters, and design scenes~\cite{dialoguedriven2017}, but seeing dialogues only as text is often inadequate to convey the multimodal sensory richness the writers conceive.
This disconnect adds uncertainty and friction to the writing process, as the writers often need to rely on their own imaginations to develop and refine dialogues~\cite{scriptviz2024}.
Moreover, writers frequently lose creative control once their scripts enter production, while artists are tasked with interpreting scripts without adequate audiovisual guidance, leading to collaboration challenges and inefficiencies~\cite{field2005screenplay, mckee1997substance}.

Writers have attempted to address these challenges by incorporating multimedia references alongside their scripts. However, this often involves a meticulous and time-consuming search for references that match their narrative intent. Prior work has explored automatic approaches to either generate~\cite{cardinal2018} or retrieve~\cite{scriptviz2024} reference visuals based on user scripts to facilitate the writing process.
However, these methods are limited in their ability to find ideal references or provide sufficient customization options.
More importantly, as suggested by the findings of our formative study with four industry professionals, 
prior methods focus solely on visuals, overlooking the fact that dialogue writing can be facilitated by incorporating multimodal audiovisual aids.
Moreover, the iterative nature of scriptwriting demands immediate feedback to refine the narrative, which traditional workflows often fail to provide, as feedback typically becomes available only during later stages, such as storyboarding or production.

Informed by these findings and prior work, we developed the interactive system \system{} (\reffig{fig:teaser}), an AI-assisted tool integrating the traditional text-only scriptwriting workflow with interactive audiovisual scene creation. As users compose scripts, \system{} automatically generates animated scenes, offering an audiovisual representation of the narrative. The system focuses on dialogue writing, where character conversations drive the plot, producing animated characters in 3D environments with expressive, synthesized speech. Users can adjust audiovisual elements and explore various interpretations of their script, facilitating iterative refinement. 

\system{} integrates text and audiovisual generation through a novel text-to-audiovisual-scene generation pipeline. The pipeline utilizes large language models (LLMs) to parse and annotate scripts, guiding the synthesis of expressive speech that captures emotional nuances. The synthesized speech then conditions the generation of character movements, ensuring alignment between vocal expressions and animations. Finally, the pipeline renders animated characters and expressive speech within interactive 3D scenes. We note that the audiovisual content \system{} generates are not meant to be finalized, production-ready content but instead function as aids for design thinking ideation. Like sketches in early design stages \cite{buxton2010sketching}, they help visualize and communicate concepts, supporting the script's communicative nature and clarifying ideas during development.

To gather initial insights into how \system{} supports dialogue writing and script development, we conducted a user study with 12 participants, including industry professionals and novices. Findings show that \system{} aids the scriptwriting process, especially during ideation and exploration, by providing audiovisual tools that support pre-production planning. Its audio feedback capabilities were noted for improving dialogue quality, leading to more realistic and emotionally engaging scripts, while its animation features offered deeper insights into scene development. Additionally, compared to a text-based baseline tool, the multimodal experience of our system enhanced writing engagement. More broadly, our work illuminates future research directions in leveraging AI to unify multimodal modalities, facilitating the content creation experience in domains traditionally confined to a single modality.

\vspace{2mm}
\noindent
This paper presents the following contributions: 
\begin{itemize}[leftmargin=1em]
\item A formative study that identifies the challenges of unimodal scriptwriting and how bridging text with audiovisual elements can enhance both the writing process and the overall video production workflow.

\item \system{}, an interactive AI tool supporting dialogue scriptwriting through interactive audiovisual generation for iterative script refinement. It features  a generative AI-based text-to-audiovisual-scene pipeline that processes script text, generating expressive speech and character animations within 3D scenes.

\item Findings from user studies highlighting how \system{}'s audiovisual generation supports ideation, exploration, and the iterative refinement process in scriptwriting.
\end{itemize}

%% file: sections/related_works.tex
\section{Related Works}
Script2Screen is related to works in writing support tools, audio storytelling, and integrative tools for text and audio-visual media.
\label{sec:related_works}

\subsection{Writing Support Tools}
Writing is one of the most essential human activities, serving to document events, convey stories, and express boundless imagination. However, the writing process presents various challenges, such as maintaining clarity, ensuring grammatical accuracy, and overcoming writer's block. To alleviate these difficulties, numerous tools have been developed to support different stages of writing, such as ideation~\cite{huang2020heteroglossia, chou2023talestream}, editing~\cite{coauthor2022, zhu2017cept}, and grammar checking~\cite{naber2003rule, cavaleri2016you}. Earlier approaches, like Soylent~\cite{soylent2015}, leveraged crowdsourcing for tasks such as shortening, proofreading, and editing documents on demand. With the increasing power of language models, AI—particularly LLMs—has been more widely integrated into writing support tools~\cite{talebrush2022, dramatron2023, coauthor2022, abscribe2024, lin2024rambler, chung2024patchview, yuan2022wordcraft, jakesch2023co}. For example, TaleBrush~\cite{talebrush2022} offers a sketch-based interface for generating stories; Dramatron~\cite{dramatron2023} generates screenplays and theater scripts based on user-provided prompts; and Coauthor~\cite{coauthor2022} presents a dataset showcasing the generative capabilities of LLMs for interactive writing. ABScribe~\cite{abscribe2024} supports the rapid exploration and organization of multiple writing variations in human-AI co-writing. Cardinal~\cite{cardinal2018} visualizes interactions between story characters and generates 3D animations to facilitate scriptwriting, while ScriptViz~\cite{scriptviz2024} queries reference images from a large movie database.

\subsection{Audio Storytelling}
Audio storytelling enriches narratives through sound, with speech as its core element. Traditional approaches to creating audio stories involve both recording \cite{rubin2015capture, seetharaman2019voiceassist, davis2003active} and editing \cite{pavel2020rescribe, venkataramani2017autodub, rubin2014generating}. Tools like Narration Coach \cite{rubin2015capture} and Voice Assist \cite{seetharaman2019voiceassist} support users in producing expressive voiceovers, addressing challenges faced by creators who may lack effective recording skills. In the editing phase, early tools like Audacity\footnote{\url{https://www.audacityteam.org/}} employed waveform-based interfaces. Recent advances, however, have introduced content-based tools, such as text-based editing systems, which allow users to operate with natural language semantics to edit speech \cite{shin2016dynamic, rubin2013content, wang2022record, morrison2021contextawareprosodycorrectiontextbased, jin2017voco, tan2021editspeech}. This shift enables editors to concentrate on storytelling rather than manipulating raw audio waveforms. As generative AI capabilities continue to evolve, new workflows for creating audio storytelling experiences using speech synthesis \cite{ren2019fastspeech, wang2017tacotron, kharitonov2023speak} have emerged. For instance, NewsPod \cite{laban2022newspod} synthesizes podcasts from news articles and allows users to interactively pose questions and receive responses relevant to the content. Similarly, NotebookLM\footnote{\url{https://notebooklm.google.com/}} \cite{huffman2024enhancing} generates two-party conversations from documents. \system{} utilizes text-to-speech synthesis for storytelling but extends it into a text-to-audiovisual scene generation pipeline, where the input for speech synthesis is the user's script, annotated by an LLM, and the output serves as a condition to guide character animation generation.

\subsection{Integrated Systems for Language and Audio-Visual Media}
Language is a uniquely powerful medium due to its descriptive nature, enabling the creation and organization of other modalities, such as visual and auditory content.
Systems like Promptify~\cite{brade2023promptify} and PromptCharm~\cite{wang2024promptcharm} translate user prompts into images.
Crosspower~\cite{xia2020crosspower} transforms linguistic structures from scripts or outlines into graphical elements and visual effects.
Crosscast~\cite{xia2020crosscast} extracts information from podcast transcripts and generates relevant visuals, such as those for travel podcasts. 
Leake et al.~\cite{concreteness2022} employed the notion of ``word concreteness'' to generate audiovisual slides from text articles.
Other recent works have leveraged generative AI to further enhance the integration between language and other modalities.
For example, audio-driven animation leverages speech \cite{Ao2022RhythmicGesticulator} or music \cite{choreomaster2021} to control visual animations.
Generative Disco~\cite{liu2023generative} allows users to create music visualizations from textual inputs, merging language with audio-visual outputs.
Beyond language-to-media generation, human-in-the-loop systems increasingly support visual storytelling through sketching, compositing, and timeline-based interfaces. SketchStudio~\cite{kim2018sketchstudio} and Layered 3D~\cite{ma2022layered3d} let creators prototype animations via 2D sketches and motion layering, encouraging narrative experimentation. Griffith~\cite{kato2024griffith} and CollageVis~\cite{jo2024collagevis} extend this by enabling collaborative storyboarding and real-time shot planning. These tools exemplify a shift toward accessible, expressive authoring environments that scaffold cinematic structure.

\begin{figure}[t]
    \centering
    \begin{overpic}[width=0.6\linewidth]{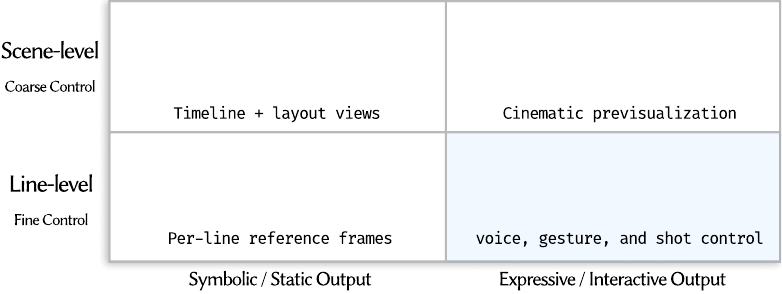}
      \put(23,28){\small\textbf{CARDINAL}~\shortcite{cardinal2018}}
      \put(72,28){\small\textbf{VDS}~\shortcite{rao2023vds}}
      \put(25,12){\small\textbf{ScriptViz}~\shortcite{scriptviz2024}}
      \put(63,12){\small\textbf{Script2Screen (Ours)}}
    \end{overpic}
    \caption{ Design space of audiovisual storytelling systems positioned by granularity of authoring control and expressiveness of output. We highlight one representative system per quadrant to reflect the most recent advances in each design space region. While prior tools focus on symbolic visualization or scene-level previsualization, \system{} uniquely supports fine-grained, expressive authoring across voice, gesture, and shot design.}
    \label{fig:design_space}
  \end{figure}

Building on these foundations, we propose a novel system that supports the scriptwriting process by integrating the textual representation of a script with its potential realization in audio and visual forms (\reffig{fig:design_space}).

%% file: sections/formative_studies.tex
\section{Formative Studies}
\label{sec:formative_studies}

To gain deeper insights into the challenges associated with the disconnection between scriptwriting and downstream audiovisual production in both modalities and work division, we conducted semi-structured formative interviews with industry professionals from the media production sector. The interviews, each lasting approximately one hour, were conducted remotely via Zoom, and participants were compensated CAD\$30 for their time. We recruited four domain experts for this study:

\begin{table*}[ht]
\centering
\rowcolors{2}{white}{lightgray} 
\caption{Participants, their roles, years of experience (YoE), and production phase primarily interested in.}
\vspace{-1mm}
\begin{tabular}{|c|c|c|c|}
\hline
\textbf{Participant} & \textbf{Role} & \textbf{YoE} & \textbf{Phase} \\ \hline
P1 & Professor at a top-ranked U.S. film school & 10+ & Pre-production \\ \hline
P2 & Screenwriter specializing in writing for games and TV shows & 1-3 & Production \\ \hline
P3 & Novelist working on the screenplay adaptation of their own book & 4-7 & Pre-production \\ \hline
P4 & Technical animator with expertise in both games and movies & 1-3 & Production \\ \hline
\end{tabular}
\end{table*}

All participants hold at least a master's degree in their respective disciplines. Below, we summarize the key insights gathered from the interviews.

\subsection{Gap between Scriptwriting and Audiovisual Production}

Participants highlighted a significant gap between scriptwriting and the subsequent stages of audiovisual production from two distinct perspectives. From the writers' perspective, P1 noted that many screenwriters often lack the visual skills necessary to convey their ideas visually, making them dependent on others and leading to a sense of lost creative control.
P1 explained, \user{"Most writers aren't fine artists… They're focused on the words—the words on the page are everything for them. It's almost like they're giving up a child for adoption."}
This sense of relinquished ownership becomes particularly evident when scripts move into production, where control is frequently transferred to visual storytellers, such as directors.

On the other hand, the visual artists' perspective reveals another dimension of this disconnect. P4, a technical animator responsible for visualizing scripts through animation, expressed the challenges of not receiving visual input from writers.
\begin{wrapfigure}{r}{0.4\columnwidth}
    \hspace{-10pt}
    \includegraphics[width=0.4\columnwidth]{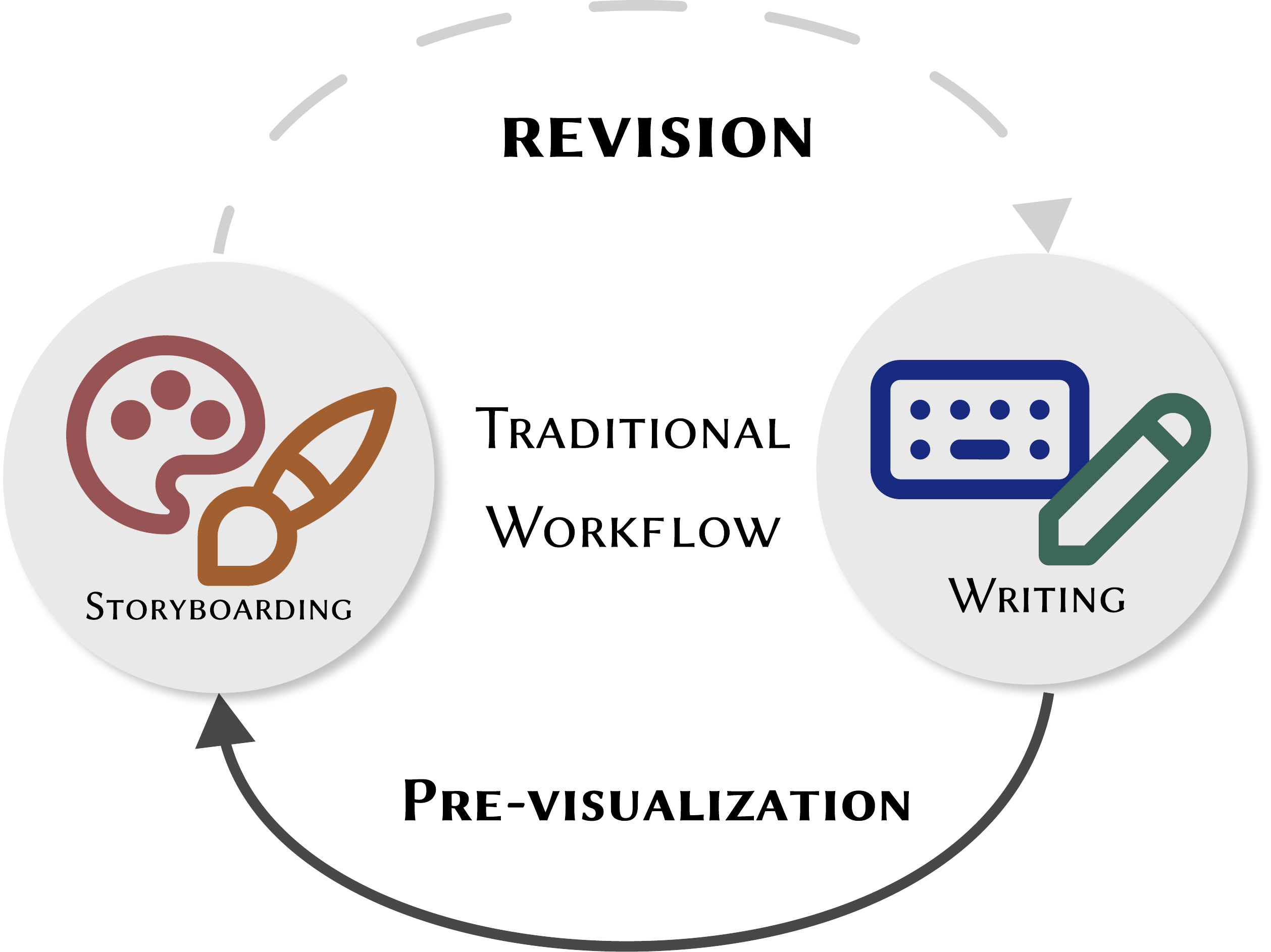}
    \hspace{-10pt}
\end{wrapfigure}
P4 highlighted that when writers do not provide clear visual references, it becomes the responsibility of animators or directors to interpret and fill in these gaps:
\user{"It would be nice if, in pre-production, they could provide reference videos of what they want things to look like, because otherwise, I have to be the one to look for them."} This reliance on production team members to make creative decisions adds a layer of complexity. While writers may feel they lose control over their work, animators and directors may feel burdened with interpreting and visualizing the script in a way that aligns with the writer's vision. This disconnect often results in discrepancies between the writer's original intent and the final product.

\subsection{Audiovisual Media as a Tool for Iterative Script Refinement} \label{sec}

Participants also identified the role of audiovisual production in refining scripts. P2 explained that storyboarding—a visual step following scriptwriting—often reveals issues and prompts revisions: \user{"Once you start storyboarding, you realize something and often go back to the script to revise."} This illustrates how visualizing scripts can expose problems or inspire new creative directions. However, when scriptwriting and audiovisual production are handled separately, this iterative process can increase communication challenges and extend production timelines. As such, P3 expressed a desire for integrated tools that provide real-time feedback, noting: \user{"It would be cool to see characters talking as you're writing. Like you're writing some script, and you can see the characters talking on the right, saying exactly what you're writing."} These insights highlight a need for tools that allow writers to see their scripts come to life as they work, helping bridge the gap between the writer's intent and the final audiovisual outcome.

%% file: sections/design_objectives.tex
\section{Design Goals}
\label{sec:design_objectives}
Our formative interviews emphasize the need to bridge the gap between scriptwriting and its audiovisual realization while supporting iterative script refinement. Informed by these findings and established human-AI interaction guidelines \cite{amershi2019}, we formulate the following design goals to guide the development of \system{}.

\vspace{6pt} \noindent\textbf{DG 1. Automating Audiovisual Scene Generation from Script Text}.
\system{} should automatically generate audiovisual representations of the scripts users write, bridging the gap between textual and audiovisual content \cite{concreteness2022}.

\vspace{6pt} \noindent\textbf{DG 2. Supporting Iterative Scriptwriting with Immediate Feedback}. \system{} should enable users to iteratively refine their scripts based on immediate feedback \cite{epstein2002immediate, wang2021soloist}, allowing for continuous improvement and alignment between the written and visual components.

\vspace{6pt} \noindent\textbf{DG 3. Providing Flexible Controls for Creative Exploration}. Beyond producing a single generative output, \system{} should offer flexible user controls to fine-tune and explore various audiovisual variations \cite{brade2023promptify, 10.1145/3173574.3173943}, encouraging creative exploration and inspiration.

\vspace{6pt} \noindent\textbf{DG 4. Promoting Transparency in AI Generation}. To mitigate the opacity typical of end-to-end generative AI models, \system{} should provide transparency into its AI pipeline \cite{larsson2020transparency, dwivedi2023explainable}, giving users insight into the generation process and improving their understanding of the system's outputs.

%% file: sections/user_interface.tex
\section{The Script2Screen User Interface}
\label{sec:user_interface}
Guided by the design goals, we developed \system{}, an interactive AI tool that provides iterative audiovisual generation to support dialogue scriptwriting. The key feature is that as users write the script, the system automatically generates audiovisual scenes featuring animated characters and expressive speech. Additionally, users have flexible controls to guide the generative pipeline, enabling creative exploration. In this section, we present the user interface of \system{} and its utility. 

\subsection{Script Editor}
\label{sec:script_editor}
The Script Editor, as shown in \reffig{fig:teaser}.B in UI and \reffig{fig:pipeline} in pipeline, occupies the central portion of the interface and is designed to allow users to write and edit their script in a familiar text-editing environment. The editor supports a range of text formatting options (bold, italics, headers, bullet points, etc.), similar to a traditional rich-text editor. This section prominently displays the current scene's script with character dialogue and actions. A “Generate” button below the text box lets users trigger the system to generate corresponding animations and visualizations for the written script (\dgone{}).

\subsubsection{Generation Button}
\label{sec:generation_button}
\begin{figure}
    \centering
    \includegraphics[width=1.0\columnwidth]{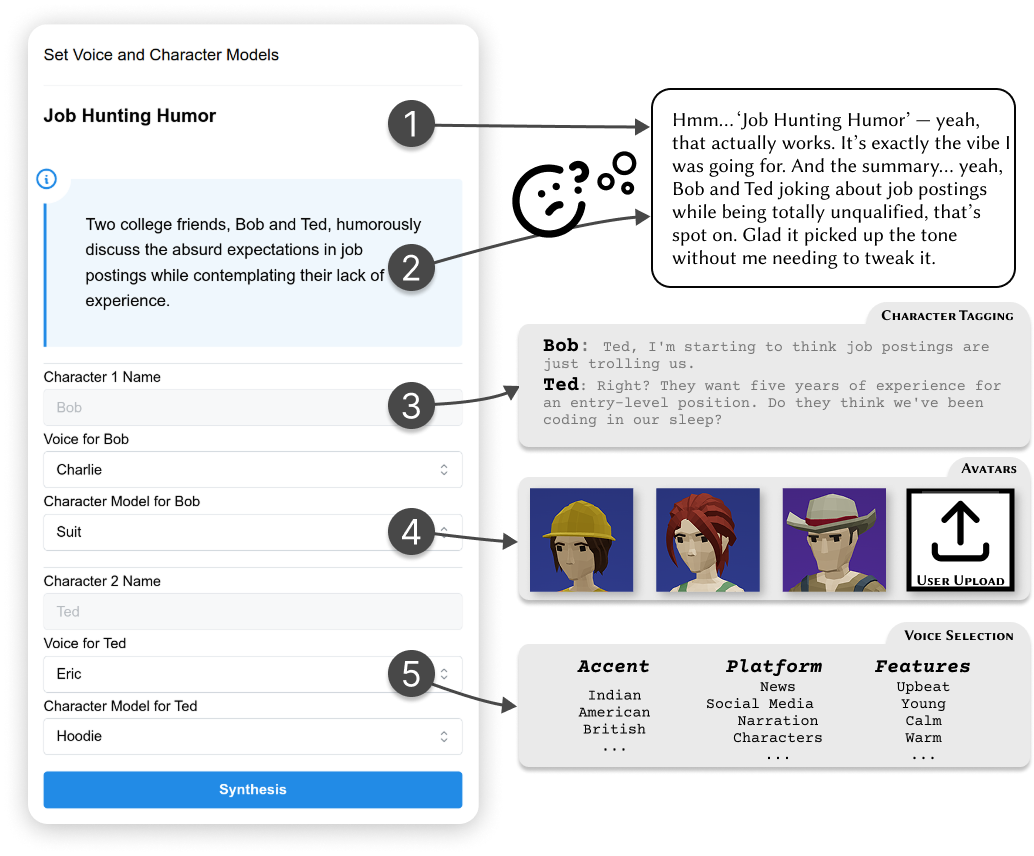}
    \caption{Before generating the animation shown in \reffig{fig:teaser}.B, the system presents a preview stage where users review an automatically generated (1) title and (2) logline summarizing their script. They can then customize the scene by (3) selecting character names, (4) choosing from a wide range of voice options and (5) visual models. Once satisfied, users click Synthesis to begin the animation process, as detailed in \refsec{sec:backend_system}.}
    \label{fig:generate_modal}
\end{figure}

On click of the “Generate” button, the system will present a pop-up window containing a \textit{title} (\reffig{fig:generate_modal}.1) and \textit{logline} (\reffig{fig:generate_modal}.2) automatically generated for the scene.  The system also provides options for users to customize the scene by selecting character voices and models (\reffig{fig:generate_modal}.3). The modal offers an intuitive interface for customizing the characters' voices and appearances before generating the final animation, ensuring that the output aligns with the desired tone and aesthetic (\dgthree{}). Below, we expand on the details of each.

\paragraph{Scene Title} A brief phrase summarizing the scene, generated by the system based on the script for a quick overview of its content.

\paragraph{Scene Logline} Displayed at the top of the pop-up window, this concise description summarizes the dialogue's context. For example, in \reffig{fig:generate_modal}, the logline captures the humorous exchange between Bob and Ted about job postings from the sample script in \reffig{fig:teaser}.B.

\paragraph{Character Voice \& Model Selection} Users can customize the characters' voices and appearances by selecting from a list of available models and voice options, each offering distinct styles and tones. At the bottom of the window, a \textit{Synthesis} button allows users to generate the scene based on their selections. When clicked, the system applies the chosen voice and character model to synthesize the animation. Upon click, the backend system (\refsec{sec:backend_system}) processes the user's selections and generates the final animation, which then populates the \textit{Timeline View} (\refsec{sec:timeline_view}) and \textit{Scene Preview} (\refsec{sec:scene_preview}) UI components.

\subsection{Timeline View}
\label{sec:timeline_view}
The Timeline View, located at the bottom of the interface (\reffig{fig:teaser}.E), serves as a parsed and linear representation of the script's dialogue and actions. Each dialogue line is displayed as an individual card (or 'blob') representing a single piece of dialogue, enriched with metadata such as emotion, generated through the LLM's analysis of the script. The metadata is clearly presented to provide transparency regarding the AI models' predictions (\dgfour{}).

\begin{figure}[h]
    \centering
    \includegraphics[width=1.0\columnwidth]{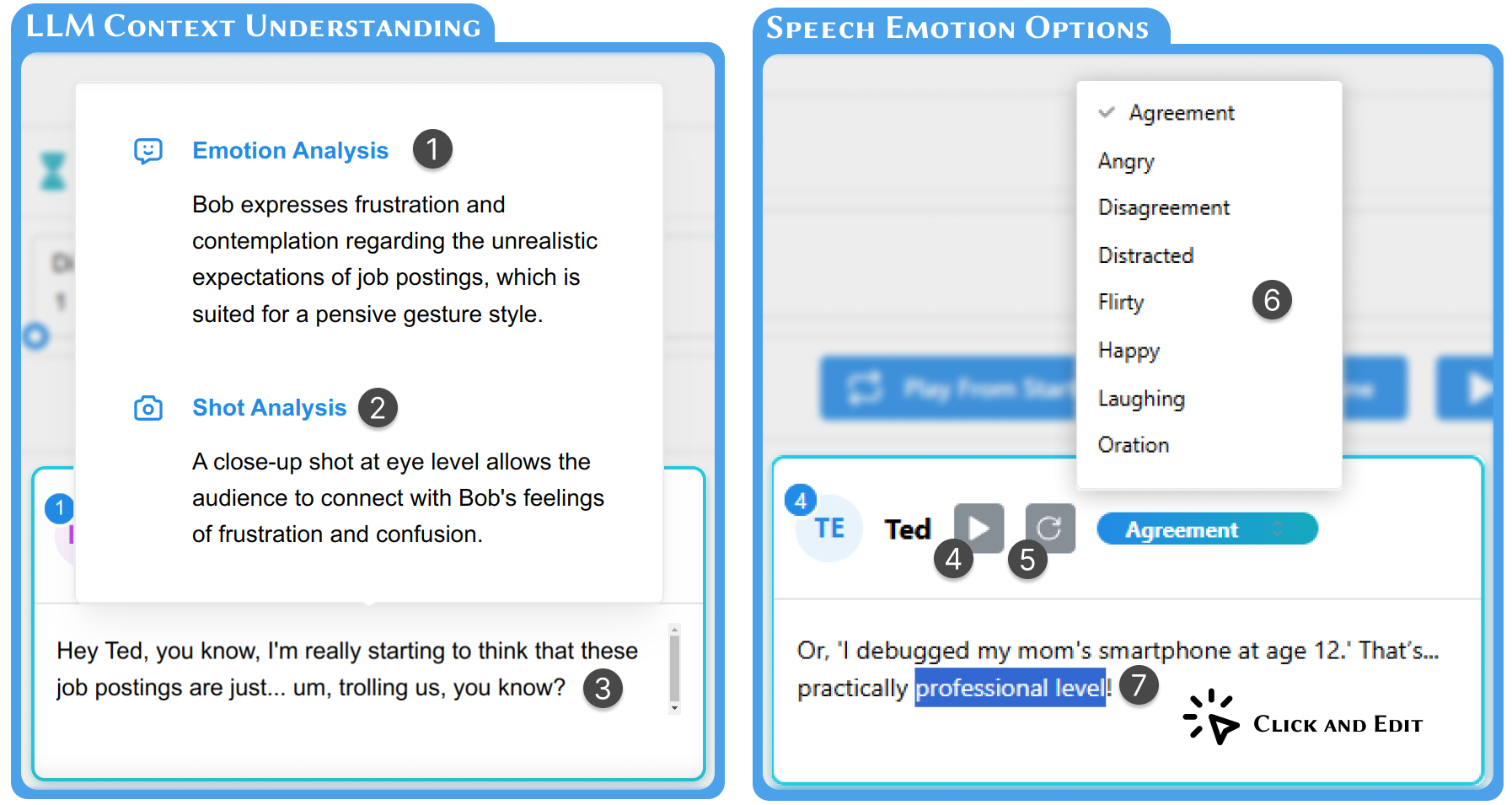}
    \caption{In each dialogue card, when the user hovers over the text area, the system highlights the (1) emotion analysis and the (2) shot analysis to explain the rationale behind the generated animations. Users can (3) directly edit the speech content like a text editor, (4) play the audio, (5) reset it, and (6) choose from various speech emotion options such as "Agreement" or "Flirty" to guide vocal tone. The text bubble (7) supports click-and-edit functionality for seamless revision.}
    \label{fig:timeline_view}
\end{figure}

\vspace{2mm}
\noindent
Key features include:
\begin{itemize}[leftmargin=1em]
    \item \textbf{Editable Text Window:} Each dialogue card contains an editable text window that shows the LLM-enhanced version of the original script (\reffig{fig:timeline_view}.3). Users can adjust the text directly, incorporating filler words, tone adjustments, and conversational nuances introduced by the LLM (\dgthree{}).
    \item \textbf{Speaker and Dialogue Information:} Each card clearly displays the dialogue ID, the character delivering the line, and an emotion/style tag that represents how the line should be delivered (e.g., "Happy," "Angry," "Pensive"). 
    \item \textbf{Interactive Controls:} Every dialogue card includes two buttons: a "play" button to preview how the line sounds with the current audio-visual setup and a "regenerate" button that reprocesses the line using the current set emotion/style. This allows users to quickly iterate and refine their scene's audio delivery (\dgthree{}).
    \item \textbf{Emotion Selection:} By clicking on the emotion/style tag, users can select from a predefined list of emotions/styles to adjust how the dialogue should be delivered. This real-time adjustment feature empowers writers to experiment with different delivery styles and immediately hear how those changes affect the performance. (\dgthree{})
\end{itemize}

The Timeline View offers a comprehensive, interactive way to engage with the script's progression, enabling users to navigate, modify, and refine the delivery of each line to create an engaging and contextually rich animated scene.

\subsection{Scene Preview}
\label{sec:scene_preview}
The Scene Preview panel, positioned to the right of the script editor (\reffig{fig:teaser}.D), displays an interactive 3D representation of the generated scene based on the written script (\dgone{}). This panel offers users the ability to gauge how their script translates into visual storytelling, serving as a tool for writer-directors to experiment with camera work and character blocking. The rendered scene features low-poly visuals with basic lighting and cartoonish 3D models.
It is not intended to be photorealistic or production-ready.
Instead, we designed it as an animation mockup to assist the scriptwriting process, similar to the role of rough sketches for visual designers~\cite{buxton2010sketching}.

\noindent
Users can customize various aspects of the scene (\dgthree), including:
\begin{itemize}[leftmargin=1em]
    \item \textbf{Camera Controls:} Users can adjust the camera shot (e.g., "Medium shot," "Close-up"), camera angle (e.g., "Eye level," "High angle"), and lock the camera to stick with the system's recommended settings for the scene. This feature allows writers to see how their script might be visualized from different perspectives.
    \item \textbf{Background Adjustment:} 
    The system uses a High Dynamic Range Image (HDRI) as a global background, allowing for lighting and reflections based on the camera’s position and angle. Users can adjust the background’s blur and intensity settings, creating an illusion of depth and focus that simulates a real-world environment. Rather than rendering a full 3D scene, which could shift attention to unnecessary visual details, the approach intentionally provides a “wireframe” feel. This minimalist representation supports the creative process by keeping users focused on the core elements without distraction.
    \item \textbf{Pan, Zoom, and Rotate:} Users have the ability to rotate, zoom, and pan the camera using intuitive controls. For example, dragging with the right mouse button rotates the scene, the scroll wheel zooms in and out, and holding the middle mouse button pans across the scene. These controls provide a degree of freedom for users to explore their scene from various angles, giving them the flexibility to direct how the scene should appear visually.
\end{itemize}

%% file: sections/backend.tex
\begin{figure*}[t]
    \centering

\includegraphics[width=\linewidth]{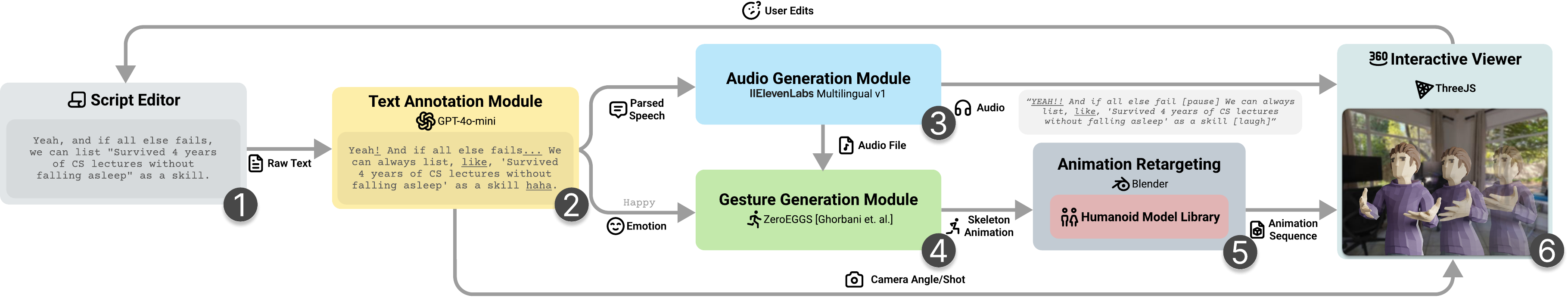}
    \caption{Overview of the \system{} text-to-audiovisual generation pipeline. The pipeline begins with the user's written script (1), which is processed by the text annotation module (2) using an LLM to parse and annotate the text, extracting information as outlined in section \refsec{sec:script_to_dialogue}. The annotated text is then sent to the audio generation module (3) to synthesize expressive speech audio. Both the speech and emotion tags are subsequently passed to the gesture generation module (4) to produce a motion capture file. This file is used in the animation retargeting module (5) to render an animated character, which is displayed on the interactive viewer (6) for user interaction. Based on the generated animation, Users can then iterate on the scriptwriting and initiate new generations as needed.}   
\label{fig:pipeline}
\end{figure*}
\section{Text-to-Audiovisual Scene Generation}
\label{sec:backend_system}
We now describe the generative model backend.
The full pipeline, consisting of script-to-dialogue (\refsec{sec:script_to_dialogue}), text-to-speech (\refsec{sec:text_to_audio}), and audio-to-gesture (\refsec{sec:audio_to_animation}) generation, ensures synchronized dialogue delivery with corresponding animations (\dgone{}). This integration results in a cohesive, interactive experience that accurately reflects the original script's intent (\dgtwo{}).
The combination of text-to-speech and gesture generation models enables \system to bridge the gap between static text and dynamic audiovisual storytelling, bringing written narratives to life in a 3D animated environment.

\subsection{Script-to-Dialogue with Large Language Model}
\label{sec:script_to_dialogue}
Upon the user clicks on the "Generate" button (\refsec{sec:generation_button}), the text annotation module (\reffig{fig:pipeline}.2) utilizes a Large Language Model (LLM) to transform user-provided text scripts into structured dialogue sequences. The system integrates the LLM through prompt-based querying, using a structured preamble and input formatting to process text efficiently.

\begin{figure*}[t]
    \centering    \includegraphics[width=1\linewidth]{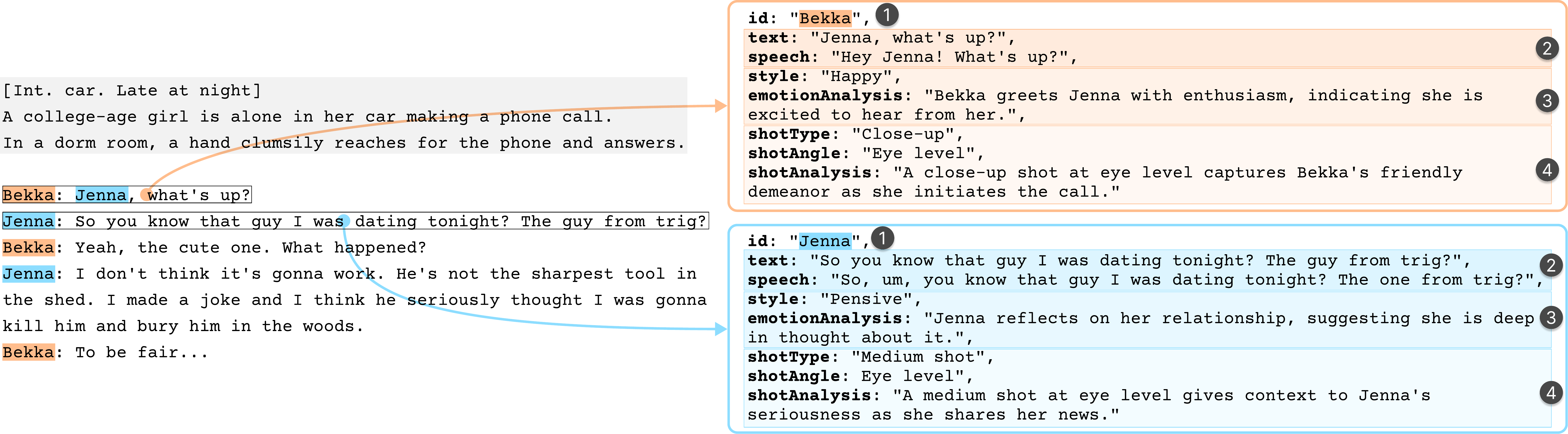}
    \caption{The text annotation module parses the script using Large Language Models (LLMs) to extract dialogue, speaker names, and other narrative elements (\refsec{sec:script_to_dialogue}). The LLM identifies characters (1), semantically parses text to verbal speech (2), assigns emotion and style labels to dialogue lines (3), and selects shot types and camera angles for each line (4). The structured output is saved as a JSON object for use in generating synchronized audio and animations.}
    \label{fig:llm_parsing}
\end{figure*}

The text annotation module performs semantic parsing: upon receiving raw text, the backend uses predefined LLM prompts (see \refsec{sec:appendix}) to summarize and process. This module parse raw text input into JSON-structured output, as visualized in \reffig{fig:llm_parsing}:

\paragraph{Character Identification} The LLM identifies individual characters based on the provided script lines, generating unique IDs for each speaker (\reffig{fig:llm_parsing}.1). This identification process is crucial for maintaining consistency and ensuring that each dialogue line is correctly attributed to the appropriate character. By establishing this structured character mapping early on, the system supports the generation of individualized audio and gesture animations in subsequent stages, preserving the integrity of character-driven narratives.

\paragraph{Semantic Understanding} The LLM converts text into speech by adding filler words or pauses (\reffig{fig:llm_parsing}.2). The system employs one-shot prompting (\refsec{sec:script_parsing_prompt}), providing a detailed example of how to parse the script into structured data. This example-based guidance ensures that the LLM accurately extracts key components such as speaker identification, dialogue, and contextual details from the script.

\paragraph{Emotion and Style Labeling} The LLM employs a chain-of-thought prompting technique to analyze each line of dialogue in the context of the script (\reffig{fig:llm_parsing}.3). This approach enables the LLM to infer the emotional tone and gesture style that best fits the character's line, selecting from a predefined list of styles (e.g., "Agreement," "Angry," "Flirty" in \refsec{sec:gesture_camera_options}). The LLM not only labels each line with a suitable gesture style but also provides an emotion analysis that explains the rationale behind its choices. For instance, when a character expresses excitement or curiosity, the LLM might select a "Happy" or "Pensive" gesture style and describe the underlying emotion as a supportive or contemplative reaction. This process ensures that the resulting animations capture the nuances of the character's emotions, adding depth and authenticity to their performance.

\paragraph{Shot Type and Camera Angle Selection} In addition to parsing dialogue, the LLM is tasked with selecting appropriate camera shots and angles for each line of dialogue, further enhancing the cinematic quality of the final output (\reffig{fig:llm_parsing}.4). The system uses chain-of-thought prompting (\refsec{sec:script_parsing_prompt}), where the LLM reasons through the context of each line to decide on the most suitable shot type (e.g., "Extreme close-up," "Medium shot") and camera angle (e.g., "Eye level," "High angle"). These decisions are accompanied by a "shot analysis," where the LLM explains why a particular shot type and angle were chosen, ensuring that each frame reflects the intended mood and visual storytelling of the scene. For example, an intense moment might be represented by a "Close-up" with a "Low angle," emphasizing the character's power or emotion in that instant.

\paragraph{Summarization and Contextual Overview} The system also employs one-shot prompting (\refsec{sec:summarization_prompt}) to generate a title and synopsis for the entire script scene. This process involves instructing the LLM to summarize the script with a caption and a concise storyline, which helps in maintaining an overarching narrative structure throughout the scene generation process (\dgfour{}). The generated title and logline serve as a context for the visual and auditory elements, ensuring that the output aligns with the intended theme and progression of the script.

\paragraph{Sequencing} Once the LLM completes the parsing, emotion labelling, and shot selection, the structured output is saved as a JSON object. This JSON contains detailed annotations for each line, including the parsed dialogue, speaker ID, gesture style, emotion analysis, shot type, and camera angle. The backend stores this structured sequence for later use in generating synchronized audio and animations. This structured format enables the system to transform raw script text into a dynamic, interactive audiovisual experience (\dgtwo{}).

The combination of zero-shot and one-shot prompting techniques, coupled with chain-of-thought reasoning, allows the LLM to parse, understand, and enrich the script content effectively. By capturing conversational nuances, character emotions, gesture styles, and cinematic elements, Script2Screen ensures that the resulting 3D animations faithfully reflect the script's narrative intent, providing an engaging and contextually accurate representation of the original text.

\subsection{Text-to-Speech with Audio Generative Model}
\label{sec:text_to_audio}
For converting text into audio, \system{} employs ElevenLabs' text-to-speech (TTS) API \cite{elevenlabs}, which generates high-quality voice output from the parsed script raw text (\reffig{fig:pipeline}.3). The backend leverages the TTS module as follows:

The system fetches a list of available voices via the ElevenLabs API, allowing users to select a specific voice for each character. This dynamic mapping ensures diverse vocal representation in the output, matching the nuances of each character's dialogue. The audio generation module ensures high-fidelity voice output, supporting a wide range of voice styles and enabling seamless integration with gesture animations. The submodule is parallelized to handle multiple lines of dialogue simultaneously, optimizing performance and reducing processing time.

\subsection{Audio-to-Animation with Gesture Generative Model}
\label{sec:audio_to_animation}
The final stage of \textit{Script2Screen}'s backend involves converting generated audio into synchronized animations using gesture generative models (\reffig{fig:pipeline}.4). This process integrates with ZeroEGGS to produce realistic body movements that correspond to spoken dialogue.

\begin{figure*}[t]
    \centering
    \includegraphics[width=1\linewidth]{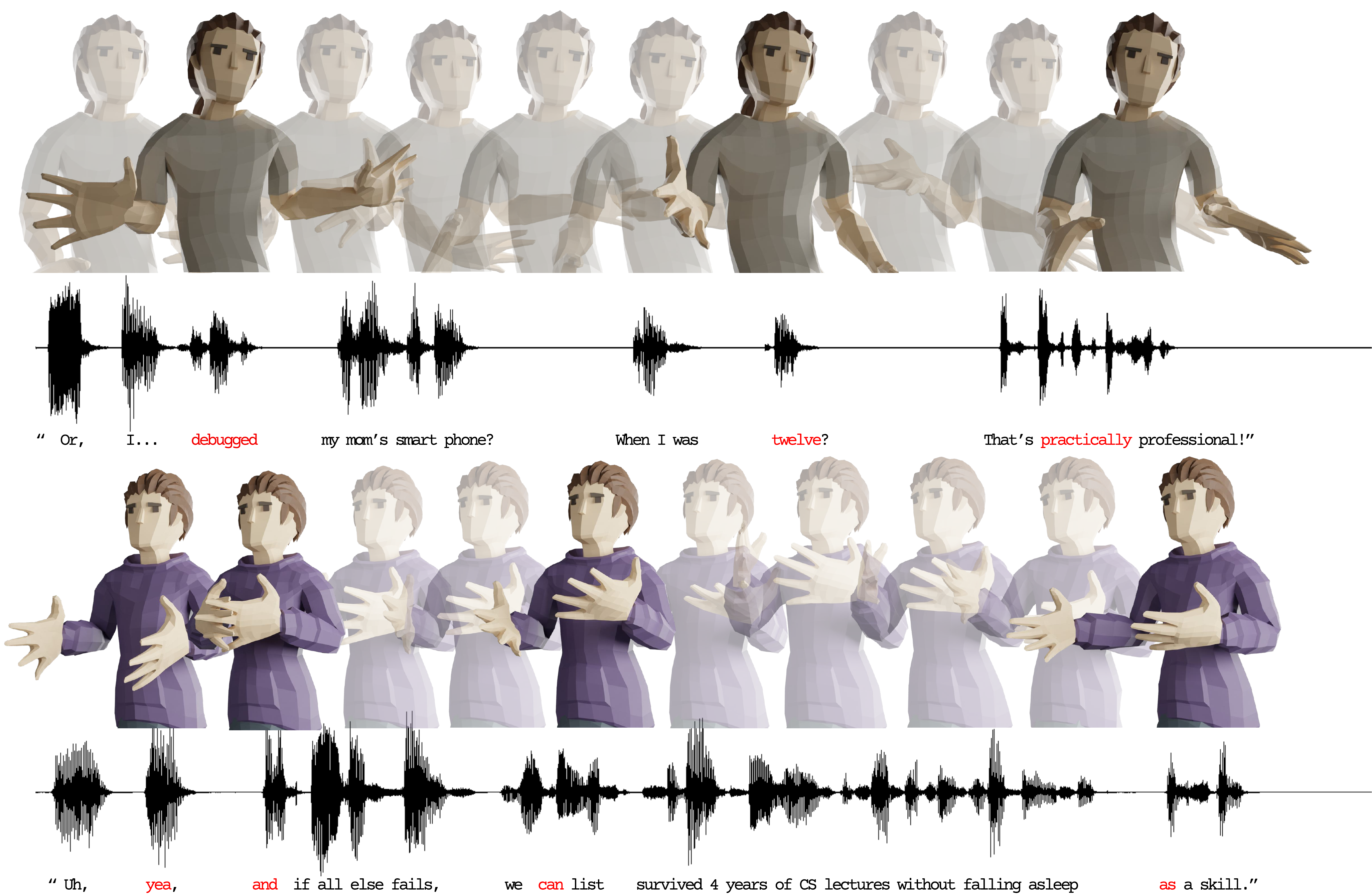}
    \caption{The system generates rhythmic character gestures synchronized with spoken dialogue, enhancing the expressiveness of the scene. The highlighted poses shown in this figure align with the text highlighted in red. This cohesive interplay between audio and animation reinforces the meaning and emotion conveyed, making the interaction more immersive and engaging for the viewer.}
    \label{fig:cospeech}
\end{figure*}

\paragraph{Gesture Generation}
The system supports various gesture styles, such as "Happy", "Sad", "Pensive", and other styles provided in \cite{zeroeggs2023}, and outputs the generated motion as motion capture files (e.g. how the character moves their hands, head, and body on skeletal joints level without the need for a 3D model) as shown in \reffig{fig:cospeech}. The complete list of supported gestures can be found in appendix \ref{sec:gesture_camera_options}.

\paragraph{Animation Conversion}
These motion capture files are further processed in the animation retargeting module (\reffig{fig:pipeline}.5) with \texttt{Blender python} to convert gesture data into animation files tailored to the \textit{user-selected} 3D character models. The system employs multiple threads for this conversion, allowing simultaneous processing of different lines of dialogue.

\reffig{fig:cospeech} presents an example of the final output from our pipeline, showcasing character animation and expressive speech. The system animates characters whose gestures naturally synchronize with spoken dialogue, highlighting keywords, phrases, and the rhythm of speech. This audiovisual experience enables writers to capture sensory elements beyond text, enriching the scriptwriting process.

\subsection{System Architecture}
We implemented the \system{} using a client-server architecture. The frontend is implemented with NextJS and React, using Mantine as the UI library. The backend is a Python Flask wrapper of the \refsec{sec:backend_system} models.
The system is deployed on a Linux machine with 64GB RAM and 1 Nvidia RTX 4090 GPU. We use OpenAI API \cite{openai2023gpt4} for \texttt{gpt-4o-mini} to process the script-to-dialogue generation in \refsec{sec:script_to_dialogue}.
The text is then audioized with the \texttt{eleven\_monolingual\_v1} model from ElevenLabs \cite{elevenlabs}.
The audio is then used to generate gestures with the ZeroEGGS \cite{zeroeggs2023} with the gesture style parsed in \refsec{sec:script_to_dialogue}.

\subsection{Pipeline Design Choice vs. End-to-End Video Generative Models}
At the time of this study, state-of-the-art audio-visual generative models such as Sora2 and Veo3 had not yet been released. Existing video generation methods also had several limitations, which motivated our pipeline-based design.

\paragraph{Generation Length.} Most end-to-end video models only support a few seconds of coherent output. In contrast, our pipeline can co-generate audio and animation of arbitrary length, limited only by the input script and compute resources. This makes it more suitable for extended dialogue scenes and interactive storytelling.

\paragraph{Iterative Control.} End-to-end generation couples visual quality with narrative control, which makes small edits difficult. For example, changing a line of text or adjusting a gesture often requires regenerating the entire video. Our modular pipeline supports iteration at different stages (script, audio, or animation), enabling a more flexible and user-driven workflow.

\paragraph{Modularity and Extensibility.} Each component of the pipeline (LLM parsing, TTS, gesture generation, and animation retargeting) is implemented as a separate module. This makes it possible to improve or replace individual parts, and to integrate new models as they become available. The modular design ensures that the system can continue to evolve beyond the current limits of end-to-end video generation.

%% file: sections/user_studies.tex
\section{User Study}
\label{sec:user_studies}
We conducted a user study to gain insights into how \system{} may assist with scriptwriting, as well as to identify limitations and explore opportunities for future advancements. Specifically, we explored how the automated generation bridged the gap between text and audiovisual content, whether the system effectively supported iterative scriptwriting, and how users perceived the flexible controls and transparency of the AI pipeline provided by \system{}. The study also gathered feedback on how our system compares to a baseline writing tool.

\subsection{Study Procedure}
The study was conducted remotely or in person, depending on participants' preferences. Each session lasted 75 minutes and was designed to compare our system, \system{}, with a baseline common writing tool, Google Docs, presented in a randomized order. 
Initially, we \textit{asked} users to choose any editor for scriptwriting.
We found that the first few expert participants all chose Google Docs over professional scriptwriting editors for \textit{short} writing tasks in our study.
They explained that professional tools such as Final Draft\footnote{\url{https://www.finaldraft.com/}} could be excessive for the tasks we designed, while Google Docs was sufficient and also allowed for live presentation of results.
Observing this trend, we decided to use Google Docs as the baseline to ensure consistency across participants (both experts and novices).

Each participant engaged in two 15-minute writing sessions (30 minutes total) to evaluate both tools. Following each session, participants completed a survey to provide their impressions of the tool. In the first part of the study, participants were instructed to write a short conversation between two people on a fixed topic, "Weekend Plan," with a suggested but flexible constraint of six sentences (three lines per person). This constraint was set to ensure efficient system performance, as \system{}’s complexity is $\mathcal{O}(n)$ based on the number of lines in the script. However, participants were encouraged to exceed this limit if they wished, allowing them to prioritize their creative preferences over strict adherence to guidelines. In the second part, participants had the opportunity to freely explore the tool until they were satisfied with their creation.

After the writing sessions, participants completed a post-study survey, which included questions adapted from the Creativity Support Index \cite{10.1145/2617588} and additional questions specific to \system{}, all rated on a 7-point Likert scale. The study concluded with a semi-structured interview, where participants shared their experiences with our tool. Each participant was compensated with a CAD \$30 cash or an Amazon gift card.

\subsection{Participants}
 We recruited six experts (P4, P7, P9, P10, P11, P12) with relevant experience in these fields, selected from online freelancing platforms and personal connections. In this study, we define experts as individuals with significant industry or academic experience in scriptwriting for visual production. To further evaluate the tool's potential in bridging the gap between writing and visualization and to understand how it might lower the barrier for scriptwriting, we also recruited six novices (P1, P2, P3, P5, P6, P8) interested in writing and visual production but without professional experience. Among them, only P5 had prior experience with scriptwriting and filming through an undergraduate film club, while the others had no experience in scriptwriting or visual production. 

\subsection{Study Results}
We now present the study results. During the study, one participant (P10) chose to withdraw; therefore, the following results are based on the data from the remaining 11 participants.
\label{sec:study_results}

\begin{figure*}[t]
    \centering
    \includegraphics[width=\linewidth]{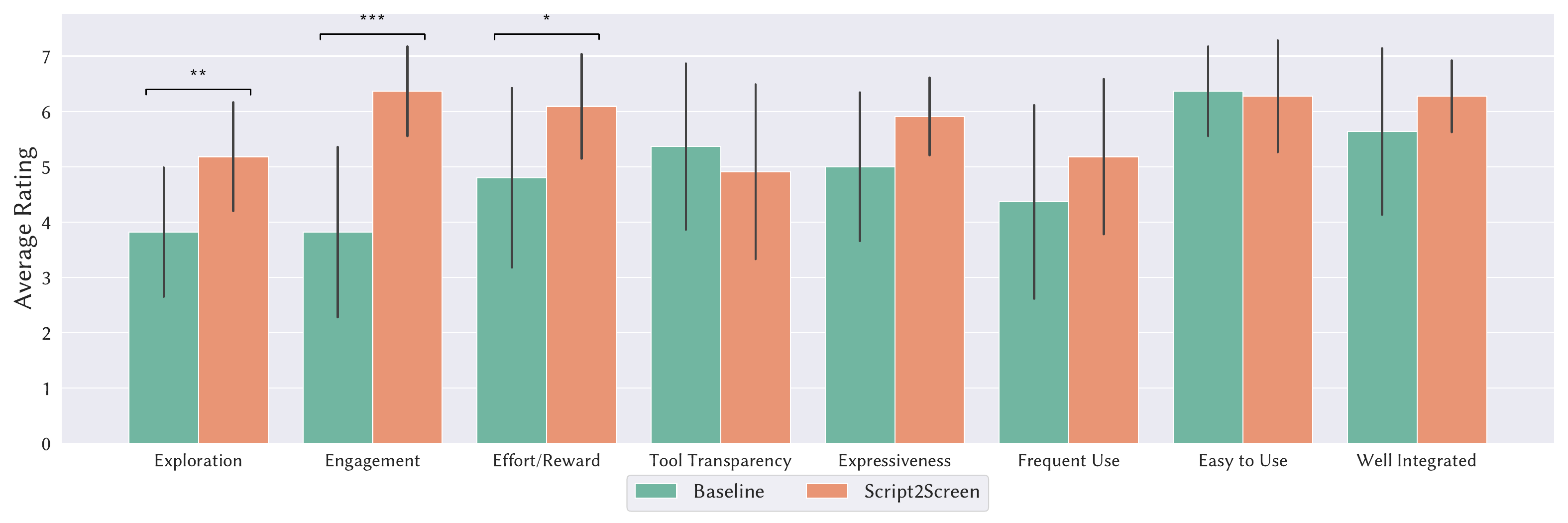}
    \caption{
    Comparison of average ratings between baseline and \system{} across the Creative Support Index. 
    Error bars represent the standard deviation. Statistical significance between the two groups is indicated with asterisks: $*$ ($p < 0.05$), $**$ ($p < 0.01$), and $***$ ($p < 0.001$). Brackets highlight significant differences, showing strong evidence of improvements in several metrics for the \system{} tool compared to the Baseline in almost all categories.}
    \label{fig:compare}
\end{figure*}

\subsubsection{Creativity Support Index Comparison with Baseline}
As shown in Figure \ref{fig:compare}, \system{} generally received positive average ratings (> 4) on the Creativity Support Index questions. Compared to the baseline condition, where users wrote in a text-based editor, our tool integrating audiovisual elements scored higher in encouraging exploration, increasing engagement, providing a sense that the reward was worth the effort, allowing for greater expressiveness, and increasing users' desire for frequent use. A t-test indicated that \system{} significantly outperformed the baseline tool in exploration (p < 0.01), engagement (p < 0.001), and perceived reward for effort (p < 0.05)

On the other hand, our tool was rated as less transparent and slightly less easy to use compared to the baseline. Users rated their confidence in using both tools equally. Although we aimed to provide transparency in the AI generation pipeline, the inherent noise and randomness in the generative process may have prevented the information from being entirely transparent to users, making the fully deterministic baseline tool appear more transparent. However, it is important to note that our tool still received a positive rating for transparency (Mean = 4.91, SD = 1.58), indicating progress toward our goals. Additionally, since this was the first time users used our tool, while they were already familiar with the baseline tool (Google Docs), the similar ratings for ease-of-use and confidence-to-use may indicate a promising level of usability for our tool.

\subsubsection{Supporting Ideation and Exploration}
As shown in both Figure \ref{fig:compare} and \ref{fig:feedback}, participants found \system{} highly effective for ideation and exploration. P1 and P2 both mentioned that they preferred using \system{} during the ideation stage. P6 further stated, \userquote{It was easy for me to explore many different options, idea designs, or outcomes without a lot of tedious repetitive interaction}, suggesting that \system{} facilitated a smooth transition from concept to execution. P7, an expert participant, noted that \system{} can help writers to \userquote{block out a page} and \userquote{gives you some really nice storyboards pretty much instantly}, highlighting how the tool's audiovisual generation capabilities accelerated the creative process. They also commented that \system{} could be helpful for pre-production. P12 echoed this sentiment, adding that our tool allowed them to concentrate on storytelling rather than technical aspects: \userquote{Compared to writing in a text editor like Final Draft, it feels more interactive, and I don't need to worry about formatting. I can just focus on the content}.

\subsubsection{Audio Feedback for Dialogue Writing}
Participants found \system{}'s text-to-audio feedback useful for crafting dialogues. P1 and P2 highlighted that hearing their scripts aloud helped refine their lines. P1 noted, \userquote{It's interesting to hear it set out, because sometimes there are things you write that only sound good in your head. But then, if you hear someone say them, it will sound awkward.} P2 mentioned that the audio made it easier \userquote{to imagine how the conversation is going on} and appreciated that it was \userquote{more emotional than...just letting the auto reading thing.} P8 added, \userquote{The audio feedback helped me see how certain phrases and words affect the tone, which was interesting.} Overall, \system{}'s auditory component facilitated organic dialogue creation, particularly benefiting novices with limited writing experience.

P9, \userquote{as someone who writes dialogue a lot}, found the audio feature valuable for externalizing the dialogue, saying, \userquote{because it's essentially taking a dialogue thing that's happening in your head and putting it on screen.} They also commented on the precision of the text-to-audio generation: \userquote{When I put in annoyed and sighs, you get that kind of essence of his character that does come across in the way that he's speaking and how he's delivering those lines.} P12 emphasized that hearing the script enhanced the emotional authenticity, stating, \userquote{The emotion helps with writing. You can tell if something feels off or not.} They also appreciated the ease of modifying emotions, adding, \userquote{I think being able to change the emotions made a big difference. It allowed me to convey the firepower of that line.} These insights underscore how the audio feedback enriched the realism and engagement in the writing process.

\subsubsection{Improved Writing and Editing Efficiency}

Many participants, especially experts, found that \system{} significantly enhanced their writing and editing efficiency.
P4 highlighted the tool's ability to streamline the editing process: \userquote{Being able to take the individual lines and being able to go into just those lines rather than having to sort of zip between the script and having to regenerate from there — I think that's helpful.} They further emphasized the flexibility and precision that \system{} provides: \userquote{The ability to edit within every blob and have control over style is helpful.} P12 mentioned that the tool accelerated their workflow: \userquote{It's definitely more speedy and convenient compared to my normal process.} They appreciated that \system{} allowed them to focus on content rather than formatting: \userquote{Compared to writing in a text editor like Final Draft, it feels more interactive, and I don't need to worry about formatting. I can just focus on the content.} Finally, P9 found the tool's summarization functionality valuable for creating log lines: \userquote{I really love that it creates a log line, by the way. I think that that's incredible, because, as a screenwriter, you study the craft of creating a log line quite a bit, and it's always a little bit difficult.}

\subsubsection{Engaging and Collaborative Writing Experience with AI}
\begin{figure}[t]
    \centering
    \includegraphics[width=1\columnwidth]{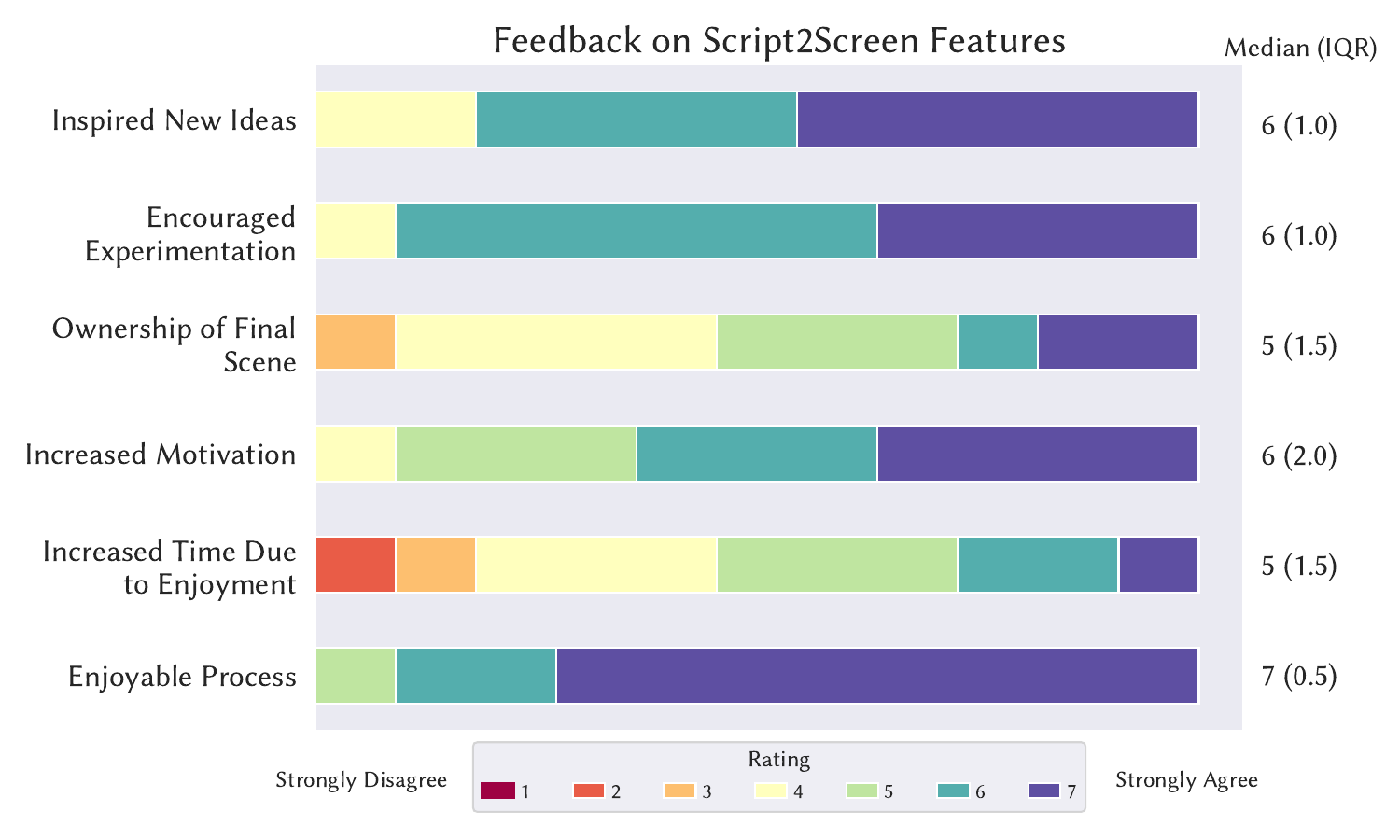}
    \caption{This bar plot illustrates the feedback from participants on various features of \system{}. Ratings range from Strongly Disagree (1) to Strongly Agree (7). Overall, the feedback for \system{} was positive, with the majority of users finding the system to be an "Enjoyable Process," particularly after a brief 15-minute session. This feature received the highest median rating of Strongly Agree (7), reflecting users' satisfaction.}
    \label{fig:feedback}
\end{figure}

We found that \system{}'s AI pipeline impacted users differently based on their level of expertise. Experts appreciated the tool's positioning of AI as a collaborator rather than a replacement, recognizing that they remained the ultimate authors of the script, even with automated generation. This is highlighted by the positive ratings of \textit{Ownership of Final Scene} in Figure \ref{fig:feedback}. In contrast, novice participants felt encouraged and motivated by the engaging writing experience \system{} offered through immediate feedback, as shown by the \textit{Increased Motivation} ratings in the same figure.

Experts such as P7 and P12 highlighted \system{}'s collaborative approach. P7 stated, \userquote{It's not trying to replace me as the writer, it's trying to help me as the writer,} emphasizing the tool's role as a supportive partner. P7, who is open to experimenting with AI-powered tools, further noted, \userquote{It's legitimately trying to be a tool for writers, which is so rare among these AI programs, so I love that.} P12 echoed this sentiment, remarking, \userquote{This tool feels like it's augmenting my creative process instead of replacing it,} reinforcing the notion of co-creation.

For novice participants, P6 appreciated the ongoing guidance, saying, \userquote{Every step of the way, I receive feedback,} and highlighting the tool's iterative nature, \userquote{The iteration process is quite good. I liked how it gave suggestions for tone and style, which helped guide my writing.} This motivated P6 to continue, stating, \userquote{It motivates me to write more because I get instant feedback on what I've written.} Similarly, P3 expressed, \userquote{The tool increased my motivation to continue,} while P8 elaborated, \userquote{It helps me see what I might have missed in my script and how I could make it more expressive... It [also] gives me hints on how I should make the script more expressive.}

\subsubsection{Feedback on Performance and Motion Richness}
Despite the overall positive feedback, participants also offered suggestions for improving \system{}. Some found it took time to adjust to the UI’s editing features. Specifically, a few novice users noted the absence of a pause feature during playback as a challenge; P2 mentioned, \userquote{I couldn't pause the video clips, which made it hard to catch everything the first time.} Participants also expressed concerns about \system{}'s performance. P1 and P3 pointed out the long processing time required to generate audiovisual scenes, with P1 stating, \userquote{I think ... speed is a huge concern.} Additionally, P3, an experienced animator, found the generated animation lacking, noting, \userquote{The character doesn’t have enough rich motion... so it’s not, you know, worth paying attention.} Overall, while our study indicated that \system{} could benefit from improvements in speed and customization features, it effectively supported the dialogue writing process by transforming plain text into multimodal audiovisual scenes.

%% file: sections/results.tex
\subsection{Example Usage of \system{}}
\label{sec:results}
In addition to the study results, we present usage examples observed during our user studies, where participants write with \system{}. We highlight two specific cases from these observations.

\subsubsection{Emotional Annotation of Script}

Providing additional context to a script can significantly enhance the \textit{speechification} process, which involves adding filler words and other elements to make speech sound more conversational and natural. This aligns the system's output more closely with the user's intent. Consider the following example script:
\begin{quote}
    \ttfamily
    Weekend plans - planning on doing some home renos, gonna reface the stairs. Alice is Bob's mom
    
    Alice: Hey Bob you remember we're doing the stairs this weekend right? You better not have made any plans with friends.

    Bob (to himself): Oh crap. How can I get out of this. I already told Suzie I'm spending the weekend with her. 
\end{quote}
The system initially generated the following emotional annotation for \texttt{Alice}'s first line, here we show a part of the JSON output:
\begin{quote}
    \ttfamily
    \dots

    \textbf{speech:} "Hey Bob! So, um, you remember we're doing the stairs this weekend, right? You really better not have, like, made any plans with friends.",
    
    \textbf{style:} "Threatening",

    \textbf{emotionAnalysis}: "Alice is asserting her authority about the plans and warning Bob not to flake on her. The threatening style conveys firmness.",

    \dots
\end{quote}
However, there are two key issues with this annotation. First, the system incorrectly labels \texttt{Alice}'s tone as \textit{threatening}. While this might be understandable based on the wording alone, given the context of \texttt{Alice} being \texttt{Bob}'s mother, this should come across more as a gentle reminder rather than a threat.
Second, the user found the system's addition of filler words like \textit{um} and \textit{like} to be out of character. These fillers made \texttt{Alice} sound younger and more casual, which conflicted with the role of a mother. To address this, the user added more specific context in the script:
\begin{quote}
    \ttfamily
    Weekend plans - planning on doing some home renos, gonna reface the stairs. Alice is Bob's mom. Alice is NOT a valley girl. Alice is an average 50 year old lady.

    \dots
\end{quote}
With this additional context, the system produced a revised emotional annotation:
\begin{quote}
    \ttfamily
    \dots

    \textbf{speech:} "Hey, Bob! So, you remember we're tackling the stairs this weekend, right? You haven't gone and made any plans with your friends, have you?",
    
    \textbf{style:} "Neutral",

    \textbf{emotionAnalysis}: "Alice is making a comment about the weekend plans but also conveying a hint of concern that Bob might not be prioritizing the family task.",

    \dots
\end{quote}
This updated annotation satisfied the user, as it aligned more closely with the character's intended role and tone. The speech now feels more authentic, reflecting \texttt{Alice}'s age and relationship with Bob. By providing more context in the script, the system generated more accurate emotional annotations. Consequently, not only is the speech more appropriate, but the entire conversation, including the audio generated from these annotated lines, aligns more closely with the user's intention.

\subsubsection{Audio Examples from Annotated Speech}

The transition from written text to audio can significantly enhance the storytelling experience by leveraging vocal elements that are difficult to capture in print. In this analysis, we will compare the original text version of a conversation between two characters with the corresponding audio generated by \system, focusing on how vocal pauses, layering of character intentions, and tone shifts improve the narrative's emotional depth and engagement. Here is a snippet of the conversation between two characters, \texttt{Bob} and \texttt{Alice}, from an user created script:

\begin{quote}
    \ttfamily
    \dots

    Bob: Wait, what!? That's insane! If you do a user study with 10 participants it will already take up, like, 10 hours!

    Alice: Yeah and it's even worse because I need to reimplement the whole software to make it work on the users' smartphone.

    Bob: You shouldn't follow her advice, you'll never make it. I can help you to convince her to think twice.

    Alice: Oh no that would never work. You know that she has a really toxic personality right?

    Bob: Ah... you're doomed.
\end{quote}

While the written dialogue conveys the general content and structure of the conversation, it lacks some of the nuances that make spoken language more immersive. The characters' emotions, hesitations, and thought processes are harder to discern in text alone. For instance, Bob's realization and resignation are present, but not as immediately clear as they would be through spoken delivery. To bridge this gap, \system{} generates an audio version of the dialogue. For clarity, we will describe the audio using a "narrative format," where the emotions and tones of the speech lines are conveyed through detailed text descriptions:
\begin{quote}
    \ttfamily
    \dots

    "Wait, what!?" Bob was shocked. "That's totally insane! If you do a user study with, like, 10 participants, it's gonna take, um, at least 10 hours!" Bob quickly said those in a surprised tone.

    "Yeah! And it's even worse because I, um, need to reimplement the whole software to make it work on the users' smartphone?", Alice said with a hint of frustration, almost questioning the decision.

    "You really shouldn't follow her advice; you won't make it!" Bob said those in a rapid style. Bob took a quick pause, as if thinking, then said "I mean, I can help you, like, convince her to think twice."

    "Oh no, that, um, would never work!" Alice emphasized on her belief of this will 'never work'. "You know she has, like, a really toxic personality, right?"

    "Ah...", Bob said as if he suddenly gets it, but also expressed it in a unfortunate tone, "you're totally doomed." Bob further confirms.
\end{quote}
In contrast, the \system{}-generated audio adds essential vocal pauses and pacing that help the listener process the conversation more naturally. \texttt{Bob}'s pauses, such as when he's thinking aloud, create a rhythm that makes his thought process more realistic and gives the audience time to follow his emotions. Similarly, \texttt{Alice}'s line, with vocal emphasis on “never work,” layers her frustration and doubt in a way that's not as evident in text. Finally, \texttt{Bob}'s resigned tone in “Ah... you're doomed” captures the shift from realization to hopelessness, adding clarity to his emotional journey. These vocal elements make the audio version more engaging and emotionally rich than the written counterpart.

%% file: sections/discussion.tex
\section{Discussions}
\label{sec:discussion}
We have demonstrated the effectiveness of \system{} in supporting dialogue scriptwriting, while also identifying areas for future exploration. We now expand on the discussions of key themes that emerged from our work and outline promising directions for future work.

\subsection{Incorporating Cross-Modality to Aid Single-Modality Tasks} 
Scriptwriting has traditionally been a single-modal task. Yet, scriptwriters are required to consider visual and audio aspects of their writing without being able to see and hear them directly.
The nature of this dilemma is what motivated the design of 
\system{}.
As our study shows, the combination of audio and visual elements can significantly assist in tasks traditionally associated with writing (text). 
More generally, certain types of information are better articulated or can only be fully conveyed through specific modalities (e.g., spatial relationships through visuals, emotions through speech in audio).
Therefore, unifying multiple modalities within a single system proves highly beneficial. 
There is a long history of creativity tools addressing challenges in one modality through the use of another (e.g., language-based editing of video~\cite{lave2024}). 
Our work resonates with this line of approach, effectively integrating three modalities in a novel and unified way. 

\subsection{User Behaviors Influenced by Prior Experience with LLMs}
We observed two instances in our study where participants utilized additional functionality provided by the LLM powering our tool without explicit guidance from the experimenters, aligning with similar findings by Wang et al.~\cite{lave2024}. For example, P7, an expert, tested the system by first inputting a 35-page production script from their own work to evaluate the tool's summarization capabilities. Following this, P7 deliberately wrote incomplete sentences to observe how the system would respond. In contrast, P5, a Ph.D. student in computer science, wrote a short script and felt dissatisfied with a particular line. Knowing that the system was powered by an LLM, P5 intentionally crafted a messy sentence and used square brackets (``[]'') to prompt the system for a cleanup. After \system{} generated the scene, P5 expressed satisfaction, noting that the system effectively refined the sentence. These examples highlight the increasing implicit knowledge users possess about interacting with LLMs for writing tasks, even without explicit instructions.

\subsection{Iterative Editing and Transparent UI for a Sense of Control}

\system{}'s ability to enable users to iteratively refine text, combined with its UI panel that visualizes how the LLM processes raw input, provides a strong sense of control, fostering further engagement. We observed that users often start by writing a few lines and then generating a scene to visualize how these lines unfold in the audiovisual output produced by \system{}.

The ability to see and hear dialogue lines reduces the cognitive load for users, who typically have to imagine the scene and dialogue flow themselves when using traditional text-only scriptwriting tools. By generating audiovisuals directly, the system allows users to focus more on the writing itself. After reviewing the animated dialogues with speech, users either continue writing or make edits—such as adjusting tone or word choice—before repeating the process.

P4 captured this process using a "thinking onion" analogy: \userquote{It's like a thinking onion. When you're cooking and don't know what to do, you start peeling and chopping an onion so you can think. It's got that vibe of I sort of know where I'm going, so let's start with this. While it's generating, I'm listening and thinking, let's see if I can come up with more as we go.}

\subsection{Human and AI as Collaborators in Creative Tasks}

Our tool emphasizes assisting users in creative tasks rather than replacing them. Rather than generating dialogues directly from textual descriptions in a single step, we employ text processing and audiovisual feedback to help writers iteratively refine their scripts. While current LLMs have the capability to generate complete scripts, they often fail to adapt to the unique writing styles of individual writers. As P9 pointed out, \userquote{I can personally always tell when something is AI generated. It just doesn't feel like a human wrote it. Not to say it doesn't read okay. It reads fine. It's just like there's something about it where you can't really get that kind of human sort of feel to it.}

Our study's findings highlighted the importance of positioning users and AI models as collaborators in tools for creative tasks. We conclude this discussion with a compelling perspective from P7, an expert scriptwriter: \userquote{I really appreciated ... that [\system{}] isn't trying to write for me. It isn't trying to replace me as the writer. It's trying to help me as the writer, and boy, that is a minority among these programs. Almost all of them are surprised to find out writers don't want a program to write for them. That's what we do, that we don't need that. We're good at it; that's the one thing in life we can do. We don't need a computer to do it for us.}

\section{Limitations and Future Work}
\label{sec:limitations_and_future_work}
Despite the promising potential of \textit{Script2Screen}, there are several limitations and opportunities for future work that should be considered to further enhance the tool's capabilities and usability. In this section, we discuss the current limitations and propose future directions for improvement.

\subsection{Performance Improvement}
Our system parallelizes all components that can be parallelized to speed up generation (e.g., querying web APIs, generating animations, and retargeting animations). We consistently observe query times of approximately 5 seconds for scene summarization and setup, and around 10 seconds for voice-over and animation generation for scenes under ten lines. However, some users (P1 and P3) have reported that the generation speed could still be improved. One of the primary challenges is the sequential dependency between the text-to-speech module and the animation generation module, where audio output from the former is required before the latter can proceed. This sequential nature makes further speed optimizations technically difficult.

\subsection{Richer Animation}
While previous works on script-to-animation often rely on querying animations from a pre-defined library (e.g., procedural animations), our use of generative animation significantly improves the naturalness of co-speech gestures. Since humans often perceive gestures as functions of speech rhythm \cite{Ao2022RhythmicGesticulator}, it is critical that generated gestures are synchronized with the speech. However, current research on generative animation has limitations, particularly regarding inter-agent or avatar interactions, which are crucial for creating realistic animations in conversational settings. Multiple participants have mentioned that while the current animation is interesting, the lack of facial animations and interactions between characters restricts the expressiveness of the generated scenes.

\subsection{Extending Beyond Dialogue Writing}

The art of cinematography extends beyond conversations to encompass a wide variety of scene types, such as establishing shots and B-reels. While our current approach excels at generating conversational scenes with basic avatar interactions, it falls short in supporting more complex scene elements. Recent research \cite{rao2024scriptviz} explored text-based queries for real film clips from databases such as MovieNet \cite{movienet}, while earlier work \cite{cardinal2018, rao2023vds} has utilized pre-set animations and scene presets, limiting the diversity of shots available in the system's library. This lack of variety becomes a significant limitation since most "real" scripts require more than just conversational scenes. Establishing shots, B-reels, and other non-dialogue elements are essential to fully visualize a script.

While advancements in generative models continue to emerge from the visual computing community, an open question remains in HCI research: \userquote{To what extent do users need to visualize scenes while writing a script?} Some users prefer to complete large portions of their script before generating a scene, while others use the generated scenes to explore and refine ideas. Another question is how scene fidelity affects the writing experience. Although some users expressed interest in more detailed scenes, many indicated that background details were not a priority. Additionally, the need for fine control over scene parameters varies among users. While some users with experience in 3D controls found camera manipulation intuitive, others found it burdensome. Thus, there is a trade-off between providing high-fidelity scene details and maintaining ease of use. Addressing these questions will guide future development to balance these factors based on user needs and habits.

%% file: sections/conclusion.tex
\section{Conclusion}
\label{sec:conclusion}

In this paper, we introduce \system{}, a novel system designed to support dialogue scriptwriting through interactive audiovisual scene generation. The system is driven by a generative AI pipeline that first employs an LLM to interpret user scripts, synthesizing expressive speech with emotion based on this interpretation. The generated speech then informs a gesture generation model, which animates characters and renders them in a 3D environment. To facilitate the exploration of different audiovisual renditions, \system{}'s UI provides intuitive controls that allow users to guide the AI generation process and experiment with variations. Our studies with both professional writers and novice users show that \system{} streamlines the scriptwriting process by integrating text and audiovisual elements into a cohesive tool More broadly, this work contributes to the research on multimodal AI tools for creativity support and human-AI co-creation.

%% file: sections/genaidisclosure.tex
\section{GenAI Usage Disclosure}

In accordance with the ACM Policy on the Use of Generative AI in Publication, we disclose that generative AI (GenAI) tools were used both as part of the research and during the preparation of this manuscript.

\begin{itemize}
    \item \textbf{Research Tooling:} Our system integrates GenAI models directly. An \emph{audio generative model} \shortcite{elevenlabs} for audio generation, a \emph{gesture generative model} \shortcite{zeroeggs2023} for animation, and \emph{large language model} \shortcite{openai2023gpt4} served as the backbone for dispatching and analyzing natural language.
    \item \textbf{Authoring Support:} We used ChatGPT for limited assistance in (i) writing plotting code for Figures~8 and~9, (ii) identifying and summarizing related works, and (iii) improving transitions and flow in the \emph{User Study} section.  
\end{itemize}

All substantive research contributions, study design, implementation, data analysis, and final interpretations were carried out by the authors.

%% file: sections/appendix.tex
\appendix
\section{Prompt Design and Examples}
\label{sec:appendix}

In this appendix, we present the detailed prompts and examples used for suggesting gestures, camera shots, camera angles, and script summarization.

\subsection{Gesture Styles and Camera Options}
\label{sec:gesture_camera_options}
The following lists of gesture styles, camera shots, and camera angles were used to generate the desired behavior in characters and camera framing:

\begin{itemize}
    \item \textbf{Gesture Styles}:
    \begin{itemize}
        \item Agreement
        \item Angry
        \item Disagreement
        \item Distracted
        \item Flirty
        \item Happy
        \item Laughing
        \item Oration
        \item Neutral
        \item Old
        \item Pensive
        \item Relaxed
        \item Sad
        \item Sarcastic
        \item Scared
        \item Sneaky
        \item Still
        \item Threatening
        \item Tired
    \end{itemize}
    \item \textbf{Camera Shots}:
    \begin{itemize}
        \item Close-up
        \item Medium shot
        \item Long shot
    \end{itemize}
    \item \textbf{Camera Angles}:
    \begin{itemize}
        \item Eye level
        \item High angle
        \item Low angle
    \end{itemize}
\end{itemize}

Originally, we included Dutch angle and Extreme-close up in the list of camera angles and shots, but we removed them due to the complexity of the implementation and we observe a lack of fidelity in our prototype implementation with the two camera options.

\subsection{Script Parsing and Gesture Style Prompt}
\label{sec:script_parsing_prompt}
The following prompt was designed to guide the model in parsing the script into JSON format, while incorporating conversational adjustments, selecting gesture styles, and analyzing camera shots and angles.

\begin{quote}
\ttfamily

Please parse the following script into JSON format. While parsing, adjust the text to sound more natural and conversational based on the context. Additionally, select appropriate gesture styles to accompany each line of the script. 
Utilize the context to ADD new filler words, hesitations, and repetitions to the speech. 
Designate gesture styles for the style of gesture associated with the speech. 
Choose a gesture style from the following list: [List of Gesture Styles].

Explain the rationale behind the choice of gesture style for each line of the script. 
Please IGNORE non-verbal cues such as facial expressions, body language, and tone of voice, 
as well as scene description in the script.

Comment on the choice of camera shot and camera angle for each line of script. 
Choose from the following camera shots: [List of Camera Shots] and camera angles: [List of Camera Angles].

For example, given the following script:
Tom: Hello, how are you feeling after your all-nighter?
Bob: I'm doing well, thank you.

The output should be in the following JSON format:
\end{quote}
\begin{lstlisting}[language=json,firstnumber=1]
[
    {
        "id": "Tom",
        "text": "Hello, how are you feeling after your all-nighter?",
        "speech": "Hello! How are you feeling, after the all-nighter?",
        "style": "Happy",
        "emotionAnalysis": "Given the context of Bob just did an all-nighter, 
        I chose a happy style to indicate Tom is being supportive.",
        "shotType": "Medium shot",
        "shotAngle": "Eye level",
        "shotAnalysis": "A medium shot at eye level is appropriate for a casual greeting."
    },
    {
        "id": "Bob",
        "text": "I'm doing well, thank you.",
        "speech": "Uh, I'm doing well... thank you.",
        "style": "Tired",
        "emotionAnalysis": "Bob is tired from the all-nighter, 
        so I chose a tired style to indicate Bob is exhausted.",
        "shotType": "Medium shot",
        "shotAngle": "Eye level",
        "shotAnalysis": "A medium shot at eye level is appropriate for a casual greeting."
    }
]
\end{lstlisting}

\subsection{Summarization Prompt}
\label{sec:summarization_prompt}
This prompt was used to generate a brief title and synopsis of the script in JSON format.

\begin{quote}
\ttfamily
Summarize the script with a caption and a synopsis in JSON format such as the following example:
\end{quote}

\begin{lstlisting}[language=json,firstnumber=1]
{
    "title": "Greeting After an All-Nighter",
    "synopsis": "Two friends greet each other after an all-nighter."
}
\end{lstlisting}

\subsection{Regeneration Prompt}
\label{sec:regeneration_prompt}
This prompt was used to update the speech, emotion analysis, and shot analysis based on a given gesture style.

\begin{quote}
\ttfamily
Please update the fields "speech", "emotionAnalysis", and "shotAnalysis" in the provided JSON 
to reflect the specified emotion, [gesture style], based on the context of the following script:

[user input script]
\end{quote}